\documentclass[checkin,showpacs,aps,prl,noshowkeys]{revtex4}
\usepackage{threeparttable}
\usepackage{amsmath}
\usepackage{lipsum}
\usepackage{multirow}
\usepackage{booktabs}
\usepackage{graphicx,color}
\usepackage{graphicx}
\usepackage{subfigure}
\usepackage{amsmath}

\newcommand{\ba}{\begin{eqnarray}}
\newcommand{\ea}{\end{eqnarray}}

\newcommand{\be}{\begin{equation}}
\newcommand{\ee}{\end{equation}}

\definecolor{pink}{rgb}{1,0.18,1.0}
\def\prl{{ Phys. Rev. Lett. }}

\def\apl{{ Appl. Phys. Lett. }}
\def\prb{{ Phys. Rev. B }}

\def\nat{{ Nature }}
\def\sci{{ Science }}

\def\jap{{J. Appl. Phys. }}

\def\np{{Nat. Phys. }}

\def\nl{{Nano Lett. }}
\def\nn{{Nat. Nanotech. }}

\def\nc{{Nat. Commun. }}

\def\sci{{Science }}

\def\cms{{Comput. Mater. Sci. }}

\begin{document}

\title{Orientation and strain modulated electronic structures in puckered arsenene nanoribbons}

\author{Z. Y. Zhang}
\affiliation{Key Laboratory for Magnetism and Magnetic Materials of
 the Ministry of Education, Lanzhou University, Lanzhou 730000, China}

\author{H. N. Cao}
\affiliation{Center for Computational Science, Korea Institute of
  Science and Technology, Seoul, 136791, Korea} 

\author{J. C. Zhang}
\affiliation{Key Laboratory for Magnetism and Magnetic Materials of
 the Ministry of Education, Lanzhou University, Lanzhou 730000, China}

\author{Y. H. Wang}
\affiliation{Key Laboratory for Magnetism and Magnetic Materials of
 the Ministry of Education, Lanzhou University, Lanzhou 730000, China}

\author{D. S. Xue}
\affiliation{Key Laboratory for Magnetism and Magnetic Materials of
 the Ministry of Education, Lanzhou University, Lanzhou 730000, China}

\author{M. S. Si$^{*}$}
\affiliation{Key Laboratory for Magnetism and Magnetic Materials of
 the Ministry of Education, Lanzhou University, Lanzhou 730000, China}

\date{\today}

\begin{abstract}
Orthorhombic arsenene was recently predicted as an 
indirect bandgap semiconductor.
Here, we demonstrate that nanostructuring arsenene into
nanoribbons can successfully transform the bandgap to 
be direct. It is found that direct bandgaps hold for narrow armchair but
 wide zigzag nanoribbons, which is dominated by the competition 
 between the in-plane and out-of-plane bondings.
 Moreover, straining the nanoribbons
 also induces a direct bandgap and simultaneously modulates effectively the
 transport property. The gap energy is largely enhanced by applying 
tensile strains to the armchair structures. In the zigzag ones, 
a tensile strain makes the effective mass 
of holes much higher while a compressive strain cause it much lower 
than that of electrons. Our results are crutial to understand
 and engineer the electronic properties of two dimensional materials
 beyond the planar ones like graphene.

\end{abstract}



\maketitle

\section{I. Introduction} 

Graphene, an ideal two-dimensional (2D) material, has charmed materials
researchers with its peculiar electronic properties that allow
electrons to move freely within its surface at high speed 
 \cite{ks-novoselov, mi-katsnelson, dl-miller, l-liao}.
 But it lacks a natural bandgap, which is a mandatory feature to control the
electric flow on and off. This largely undermines graphene's usefulness as a
replacement for the mainstream semiconductors in optoelectronics. 
Just recently, few-layer arsenic was demonstrated to be an alternative 2D
semiconductor \cite {z-zhang, negative, c-kamal}. It has a sizable
bandgap (around 1 eV) and simultaneously retains
 a considerable carrier mobility  
(several thousand square centimeters per volt-second).
  From the point of view of applications, 
nanodevices are expected to be fabricated by using a single atomic
layer, wherein electron are strictly confined within the 
surface. In pricinple, this allows the fastest flow of carriers across the
surface, like graphene. Unfortunately, the 
intrinsic bandgap of monolayer arsenic (arsenene) is 
indirect, which is not desired for realistic device applications.

Generally, fabricating nanoribbons of materials are 
 expected to have large changes in electronic properties 
due to the edge and nanoscale size effect
 \cite{v-barone, my-han, j-xie, x-han}.  
 Along this perspective, a study on arsenene nanoribbons would be
an interesting research topic. It is expected to modulate
the electronic structures of arsenene and possibly to introduce an 
indirect-direct bandgap transition.
 On the other hand, in practical applications, especially in
 nano-devices, the involved materials are always  thin strips
 with edges. In experiments, arsenene nanoribbons
 can be obtained either by cutting mechanically exfoliated
arsenne, or by patterning epitaxially-grown arsenene.
Especially, the technique of scanning tunneling microscope (STM)
 lithography allow producing any pattern,
 width and edge shape of nanoribbons \cite{l-Tapasztd}.
 In this respect, it is timely
 and crucial to thoroughly investigate the electronic properties of
 arsenene nanoribbons to extend their use in semiconducting industry. 
 However, as far as our knowledge goes, there is no such study committed 
 up to now.

In this work, two types of hydrogen passivated arsenene nanoribbons with
 armchair (aANRs) and zigzag (zANRs) edges are considered and their
 electronic  structures are investigated by employing
 density functional theory (DFT) calculations. 
Our calculations demonstrate that indirect-direct bandgap transition
 is possible in both aANRs ad zANRs only by  changing
 the ribbon width. Most interestingly, the indirect-direct bandgap
 transition holds for narrow aANRs but wide zANRs.
 Furthermore, straining the nanoribbons can also transform
 the bandgap to be direct. At the same time,
 the  effective masses of carriers
can be largely modified under stains. All these properties make arsene 
as an appealing 2D materials for future applications in optoelectronics.

The rest of this paper is arranged as follows. In Sec. II, 
we briefly describe the method used in this work.
 Results and discussion are represented in Sec. III.
 Finally, we conclude our work in Sec. IV.

\section{II. Theoretical method} 
All the calculations in this work are performed with Vienna {\it ab initio}
 Simulation Package  (VASP) \cite{gkresse,
  gkjf}. The generalized gradient approximation (GGA) of 
Perdew-Burke-Ernzerhof (PBE) functional \cite{perdew}
 is adopted to describe the
exchange-correlation energy between electrons within the
projector augmented wave (PAW) method \cite{peb}.
 The plane wave energy cutoff is set
to 450 eV to ensure the convergence of total energy. Structural
optimizations are applied by relaxing the positions of all the atoms
until the convergence tolerance of force on each atom is less than
0.01 eV/{\AA}. Periodic boundary condition is used to simulate the 2D
infinite sheet. A vacuum space at least 10 {\AA} is used to avoid the
interaction between periodic images. The reciprocal space is sampled
by a fine grid of Gamma-centered Monkhorst-Pack's mesh 11 $\times$
1 $\times$ 1 in the Brillouin zone.

\section{III. Results and discussion}

\section*{A. Geometric properties of arsenene nanoribbons.}
By cutting  arsenene sheet perpendicular
 to the zigzag and armchair directions,
 two types nanoribbons with armchair (aANRs) and  zigzag (zANRs)
edges are respectively generated. These aANRs and zANRs are further
classified by the number of As-As rings ($N_{\rm r}$) 
contained in the structure. Wherein, each As-As ring includes six As atoms,
as illustrated in the planar view in Figs. 1(a) and 1(b).
An aANR (zANR) with $N_{\rm r}$ As-As rings is represented as 
$N_{\rm r}$-aANR ($N_{\rm r}$-zANR).
Two structures of 4-aANR and 4-zANR ($N_{\rm r}$ = 4) 
are respectively given in Figs. 1(a) and 1(b) for examples.
It should be noted that all the nanoribbons
 considered in our calculations are passivated with hydrogen.
For an evaluation of the thermal stability,
 the edge formation energy ($E_{\rm edge}$) is estimated as
\be
E_{\rm edge}=\frac{1}{2W_{\rm D}}(E_{\rm ribbon}-N_{\rm a}E_{\rm a}-\frac{N_{\rm H}}{2}E_{\rm H_{2}})
\ee
where $W_{\rm D}$ is the width of the nanoribbon, $E_{\rm ribbon}$ is the total
energy of the nanoribbon, $N_{\rm a}$ is the number of As atoms in
the nanoribbon, $E_{\rm a}$ is the energy of arsenene per atom, $N_{\rm H}$
is the number of hydrogen atoms, and $E_{\rm H_{2}}$ is the energy of a
H$_{2}$ molecule. Our calculation results  (see Fig. 1(c)) indicate that
 the formation of edges are slightly endothermic
 for both aANRs and zANRs, and 
the order of stability is zANRs $>$ aANRs. 
Of the nanoribbons, the armchair ones are more
 easily to be stretched than to be compressed,
with respect to that the total energy under stretching  is 
much lower than that under compressing (see the left scale of Fig. 1(d)). 
Stretching and compressing the zigzag ones are
revealed  to be nearly equivalent (see the right scale of Fig. 1(d)).
Additionally, a basic knowledge of the bonding feature of arsenene
is necessary to understand the electronic properties.
As illustrated in the inset of Fig. 1(c), 
 each unit cell of arsenene includes four inequivalent As atoms.
Wherein, the $r_{1}$ and $\theta_{1}$ describe mainly the in-plane bondings.
and in contrast, the  $r_{2}$ and $\theta_{2}$ represent primarily 
the out-of-plane bondings. 
Changes of these parameters will spark a huge effect 
on electronic structures, as discussed in the following parts.

\section*{B. Electronic structures of aANRs}
Usually, nanostructuring a semiconducting material 
into nanoribbons can modulate effectively the
 gap value ($E_{\rm g}$) due to the quantum
 confinement effect \cite{x-han, ad-yoffe, jw-son}.
It is true in our case.
As can be found in Fig. 2,
the amplitude of bandgap in aANRs gradually increases
 when it goes from 8-aANR
to 2-aANR. The gap value scales inversely 
with the ribbon's width of $W_{\rm D}$,
 as listed in Tab. I.
The $E_{\rm g}$  increases from 0.79 eV to 1.15 eV with $W_{\rm D}$ 
 changing from 31.7 {\AA} to 9.5  {\AA}.
Besides, it is found that the nature of the bandgap 
(i.e., direct or indirect)
largely depends on the ribbon's width. 
An indirect-direct bandgap transition
 takes place when reducing the ribbon's width.
 As shown in Figs. 2(a)-2(c), 2-aANR, 3-aANR and 4-aANR 
exhibit direct bandgaps.
Both the conduction band minimum (CBM) and
 valence band maximum (VBM) locate 
at the same crystal point $\Gamma$.
 While in 5-aANR (Fig. 2(d)), 6-aANR  (Fig. 2(e)),
 7-aANR  (Fig. 2(f)) and 8-aANR (Fig. 2(g)), the bandgaps
appear as indirect. The CBM is still located at the $\Gamma$ point
 while the VBM rises along
the $\Gamma$-X line and close to the X point.  Among the 
aANRs, direct bandgaps  are 
grasped by the narrower ones with $W_{\rm D}$ $\le$ 16.9 \AA.

Indeed, such an indirect-direct bandgap transition
reflects the structural modifications 
when  arsenene is cut into nanoribbon structures. 
From inspection of Tab. I, one should have noticed that
along with reducing the ribbon's width, the lattice constant 
along the infinite direction (denoted as `$a$' in Tab. I)
 decreases. This consequently affects the bond lengths
of $r_{1}$ and  $r_{2}$ and the bond angles of 
 $\theta_{1}$ and $\theta_{2}$.
When the ribbon's width is reduced,
$r_{1}$ increases while $r_{2}$ decreases (Fig. 4(a));
 $\theta_{1}$ increases in contrast to
$\theta_{2}$  decreasing  (Fig. 4(b)).
As it is found, the increase in one of the bond lengths (or angles)
leads to decrease in the other bond length (or angle).
At narrower width, both $r_{2}$ and $\theta_{2}$ are more decreased.
As a result, the out-of-plane bondings, mainly p$_{\rm z}$-orbital alike,
is strengthened. An decrease in energy is a natural 
consequence. This effect is clearly demonstrated from the orbital-resolved
 density of states (DOS), as displayed in the bottom panels in Fig. 2.
In the narrower structures, such as 2-aANR (Fig. 2(a)),
 3-aANR (Fig. 2(b)) and 4-aANR  (Fig. 2(c)), 
the p$_{\rm z}$ orbitals occupy a lower energy in the valence band (VB) 
and the VBM is dominated by the  p$_{\rm y}$ orbitals of in-plane bondings.
 While in the wider ones from
5-aANR to 8-aANR (see Figs. 2(d)-2(g)), 
the case is converse.  The p$_{\rm z}$ orbitals rise in energy
and predominate the VBM. Such a competition between the 
in-plane and the out-of-plane bondings manipulate the energy gap 
to be direct or indirect.

By using the nearly free electron model,
 the effective mass of holes at the
VBM ($m^{*}_{h}$) and of electrons at the CBM ($m^{*}_{e}$)
 are calculated using
$m^{*}=\hbar^{2}/(\partial^{2}E/\partial k^{2})$.
 In this work, we only concern  
the $m^{*}_{h}$ and  $m^{*}_{e}$ at the $\Gamma$ point in direct bandgaps.
We find that $m^{*}_{h}$ and $m^{*}_{e}$ are comparable in aANRs.  
In 16.96 {\AA} wide 4-aANR,
$m^{*}_{h}$ = 0.24$m_{0}$ and $m^{*}_{e}$ = 0.30$m_{0}$ ($m_{0}$ is the
static electron mass).  Nanostructuring  modifies these
values slightly, for example, in the narrowest 9.53 {\AA} wide 2-aANR,
$m^{*}_{h}$ = 0.37$m_{0}$ and $m^{*}_{e}$ = 0.45$m_{0}$.

\section*{C. Electronic structures of zANRs}
Now, we switch our attention to zANRs.
The quantum confinement effect is obviously observed  in zANRs.
 The $E_{\rm g}$ increases from
 1.06 eV to 1.87 eV by decreasing the 
  $W_{\rm D}$  from 37.7  {\AA} to  8.9  {\AA}
 (see Tab. I). 
Strikingly different from the case of aANRs, in zANRs,
the indirect-direct bandgap transition happens when increasing
 the ribbon's width of  $W_{\rm D}$.
As displayed in Figs. 3(a)-3(d), the bandgaps are 
indirect in $N_{\rm r}$-zANR with
$N_{\rm r}$ $\le$ 5 ($W_{\rm D}$  $\le$ 23.4 \AA).
 The CBM occurs at the $\Gamma$ point
while the VBM lies along the $\Gamma$-X line and close to the 
$\Gamma$ point. As the ribbon width is increased up to
 $\sim$28.2 \AA, the bandgap turns to be direct 
with both the CBM and VBM located at the $\Gamma$ point.
 Therefore, direct bandgaps hold 
for the relatively wider zigzag nanoribbons.
In contrast to that in aANRs,  the values of $m^{*}_{h}$
and $m^{*}_{e}$ in zANRs are relatively larger.
 In 6-zANR,  $m^{*}_{h}$ = 5.96$m_{0}$ and  $m^{*}_{e}$ = 5.09$m_{0}$.
While in 8-zANR, $m^{*}_{h}$ = 1.63$m_{0}$ and $m^{*}_{e}$ = 1.25$m_{0}$.

Similar with that in aANRs, the indirect-direct bandgap transition
 in zANRs is closely related to  the structural changes, as illustrated in
Figs. 4(c) and 4(d). By widening the ribbon's width,
the lattice constant along the infinite direction (represented as
`$b$' in Tab. I) increases. 
As a result, the values of $r_{2}$ and  $\theta_{2}$
 keep almost unchanged; 
 whereas, $r_{1}$ and  $\theta_{1}$ are increased.
 The increase in $r_{1}$ and  $\theta_{1}$
will weaken the in-plane covalent bondings
 and  give rise to an  energy increase correspondingly.
This is why the VBM in the wider structures
 are predominated by the in-plane p$_{\rm y}$ orbitals
 (from 6-zANR to 8-zANR, see Figs. 3(e)-3(g)).
Oppositely, the VBM is occupied by the  p$_{\rm z}$
orbitals in the narrower ones from 2-zANR to 5-zANR
 (see Figs. 3(a)-3(d)). Here also we clearly see the 
indirect-direct bandgap transition
is controlled by  a competition between the in-plane
 and out-of-plane bondings. The mechanism is the same as that 
discussed above in aANRs. It is that, contracted in-plane
 bonding parameters ($r_{1}$ and  $\theta_{1}$) 
 stabilize the p$_{\rm y}$ orbitals in the VB; 
shrinked out-of-plane bonding parameters ($r_{2}$ and  $\theta_{2}$) 
stabilize the  p$_{\rm z}$ orbitals. 
The relatively higher energy states ultimately dictate the VBM.
 We should note that these structural changes
 occur naturally at different ribbon's widths.
 Further reducing the parameters 
by external fields, such as straining, 
 will cause an energy increase, discussed as below.

Based on the above discussion, it is highly expected to achieve
direct bandgaps by manufacturing
narrow armchair or wide zigzag arsenene nanoribbons. 
Our results are insightful in consideration of the following two aspects.
First, an intrinsic direct bandgap make arsenene applicable
 in optoelectronics devices. Also, it would be of great interest for
valleytronic applications via achieving degenerate
energy valleys  at different $k$-points
  \cite{kf-mak2, t-cao, sf-wu, z-zhang2}. 
Second, the indirect-direct
bandgap transition observed in arsenene nanoribbons provides
researchers with solid  evidence
 to engineer the bandgap properties of similarly puckered structures 
or other non-planar structures. 
Future experiments can test our proposal directly.

\section*{D. Electronic structures of aANRs at strains}

To further understand the effects of structural changes and
engineer the electronic properties of the nanoribbons,
strain is applied along the infinite direction
 of each type of nanoribbons.
 The magnitude of the strains are employed up to a magnitude of $\pm$10\%,
which is physically realizable considering that it is much lower
than the theoretical ultimate strain of the  puckered structure
 \cite{c-kamal, x-peng}.  We assign the positive (negative)
  values of $\varepsilon$ for tensile (compressive) strains, respectively.
Here, we present the results for 5-aANR  and
5-zANR only but similar trends hold for other ribbon widths.

As illustrated in Fig. 5, the armchair structure
 is particularly sensitive to tensile
strains. Although 5-aANR possesses an intrinsic indirect bandgap,
a very low tensile strain of $\varepsilon$ = +1\% 
successfully induce a direct bandgap.
By contrast, a much higher
compressive strain of $\varepsilon$ = -10\% is required
to make the transition happen. Additionally,
tensile strains apparently enhance the bandgap width.
As  strain is applied
between  $\varepsilon$ = +1\% and  $\varepsilon$ = +10\%, 
the bandgap increases from 0.94 eV to 1.12 eV (see Tab. II).
In contrast, compressive strain reduces the energy gap. 
 Straining the structure at $\varepsilon$ = -1\% and  $\varepsilon$ = -10\% 
 corresponds to the gap values of 0.86eV and 0.61 eV, respectively.

An analysis of the geometric structures is necessary to 
under the band structures changes.
The analysis results are plotted in Figs. 7(a) and 7(b).
Both the bond length ($r_{1}$ and  $r_{2}$) and
 bond angle ($\theta_{1}$ and  $\theta_{2}$)
change linearly with the amount of strain.
When straining is applied from $\varepsilon$ = -10\% 
to $\varepsilon$ = +10\%, $r_{1}$ and  $\theta_{1}$ decrease
 while $r_{2}$ and  $\theta_{2}$ increase.
A closer look at the orbital-resolved DOS plots in 
Fig. 5 reveals that such structural changes makes the 
VBM is first dominated mainly by p$_{\rm x}$ and  p$_{\rm z}$ orbitals
(from $\varepsilon$ = -10\% to $\varepsilon$ = +2\%),
then by p$_{\rm y}$ and  p$_{\rm z}$ orbitals
(from $\varepsilon$ = +5\% to $\varepsilon$ = +7\%),
 and finally by only the p$_{\rm y}$ orbitals ($\varepsilon$ = +10\%).
As discussed previously,
the VBM of 5-aANR under zero strain is comprised of 
 p$_{\rm x}$ and  p$_{\rm z}$ orbitals (see Fig. 2(d)).
Stretching this armchair structure along the armchair direction
seems to be capable of stabilizing the p$_{\rm x}$ and  p$_{\rm z}$ orbitals.
Hence, the  p$_{\rm x}$ and  p$_{\rm z}$ orbitals occupy the relative lower
energy and the  p$_{\rm y}$ orbitals dominate the VBM.
Similar phenomenon was observed in
the similar structure of armchair 
puckered phosphorene nanoribbons \cite{x-han}. 
In addition, at stretching strains,
 both $m^{*}_{h}$ and  $m^{*}_{e}$ is lowered
slightly in comparison with the case of no strain (see Tab. II).
However, this does not affect
 the carrier transport property
much since the ratio $m^{*}_{h}/m^{*}_{e}$  keeps almost unchanged.

\section*{E. Electronic structures of zANRs at strains}
The calculated band structures of 5-zANR
 at various strains are presented
in Fig. 6. We can see that, at a low tensile or compressive  strain 
 of $|\varepsilon|$ $<$ 5\%,
 the bandgap remains as indirect. In a contrast,
more strengthened stretching 
($\varepsilon$ $>$ +5\%) or
compressing ($\varepsilon$ $<$ -5\%) 
make the bandgap transfer from indirect to direct.
In addition to this, in contrast to that in aANRs,
both tensile and compressive 
strains result in decreased bandgaps in 5-zANR (see Tab. II).
The gap energy decreases from 1.21eV to 0.24 eV when
straining is increased from $\varepsilon$ = 0 to $\varepsilon$ = +10\%.
The value changes from 1.21eV to 0.32 eV when
strain is applied from $\varepsilon$ = 0 to  $\varepsilon$ = -10\%.

The structural changes of 5-zANR 
at various strains are plotted in Figs. 7(c) and 7(d). 
Both the bond length ($r_{1}$ and  $r_{2}$) and
 bond angle ($\theta_{1}$ and  $\theta_{2}$)
change linearly with the amount of strain.
As the strain varies from $\varepsilon$ = -10\% 
to  $\varepsilon$ = +10\%, $r_{1}$ and  $\theta_{1}$ increase
 while  $r_{2}$ and  $\theta_{2}$ decrease. 
This consequently imposes
 an influence on the electronic states near the VBM. 
At compressive strains, the VBM is composed mainly with
the in-plane bondings of p$_{\rm x}$ orbitals (from $\varepsilon$ = -10\% 
to $\varepsilon$ = -7\%) and of p$_{\rm y}$ orbitals
 (from $\varepsilon$ = -5\% to $\varepsilon$ = -1\%).
By comparison, at all tensile strains
 from $\varepsilon$ = +1\% to $\varepsilon$ = +10\%,
 the VBM is predominated by 
the out-of-plane bondings of  p$_{\rm z}$ orbitals.
This is because that, around the optimized structures,
reducing the $r_{1}$ and $\theta_{1}$  causes an energy 
increase of the in-plane bondings of p$_{\rm x}$ and  p$_{\rm y}$
orbitals; decreasing the $r_{2}$ and $\theta_{2}$ lead to the energy 
increase of the out-of-plane bondings of p$_{\rm z}$
orbitals, sharing the same mechanism as that  in 
 strained aANRs.  In addition, straining effectively tune the
$m^{*}_{h}$ and  $m^{*}_{e}$, especially the ratio of 
$m^{*}_{h}/m^{*}_{e}$ (see Tab. II). At strains of  $\varepsilon$ $>$ 0,
 $m^{*}_{h}$ is much larger than $m^{*}_{e}$. The ratio of $m^{*}_{h}/m^{*}_{e}$
is around  3.7-6.5. In a
sharp contrast, at strains of  $\varepsilon$ $<$ 0,
 $m^{*}_{h}$ is much smaller than $m^{*}_{e}$ with
the ration of  $m^{*}_{h}/m^{*}_{e}$ = 0.1-0.2.

\section{IV. Conclusion} 
In summary, two types of arsenene nanoribbons are constructed and
modeled. An indirect-direct bandgap transition is successfully realized by
cutting narrow aANRs and wide zANRs. 
The indirect-direct bandgap transition is dominated by 
the competition between the in-plane and out-of-plane bondings.
The energy gap of these nanoribbons can be modulated over a wide range
by varying the ribbon's width.
Placing  aANRs and zANRs under strains significantly modifies the
band structures as well as the carrier transport properties.
Either stretching or compressing the nanoribbons
 with an appropriate strength can
 implement an indirect-direct bandgap
 transition in both aANRs and zANRs.
In addition, in aANRs, the energy gap can be largely enhanced
by applying tensile strains. 
In zANRs,  tensile strain results in $m^{*}_{h}$ much larger than
$m^{*}_{e}$ while compressive strain makes $m^{*}_{h}$ much smaller
than $m^{*}_{e}$. Our results demonstrate the
wide possibilities to tune the electronic properties of two dimensional puckered
structures. Further experimental investigations of these structures would be
of great interest.

\section{Acknowledgments} 
This work was supported by the National Basic Research Program of
China under Grant No. 2012CB933101 and the National Science
Foundation under Grant No. 51202099.

$^{*}$Email: sims@lzu.edu.cn

\clearpage

\begin{figure}
\includegraphics[width=16cm]{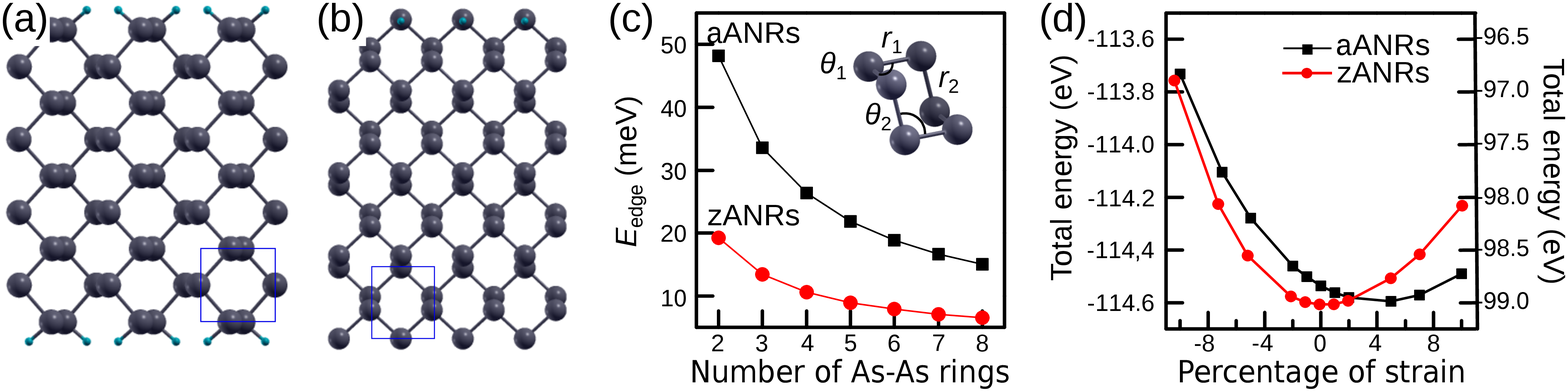}
\caption{(color online) The planar views of arsenene nanoribbons with 
edges of (a) armchair and (b) zigzag. 
The edges are passivated with hydrogen atoms
represented by small blue balls.
Both the structures contain four As-As rings in width and  
hence are denoted as (a) 4-aANR and (b) 4-zANR. 
Each As-As ring includes six As atoms,
 indicated by blue rectangles. 
 (c) The calculated  edge  formation energy ($E_{\rm edge}$)
 for aANRs (black cubics) and zANRs (red circles) depending on the
  number of As-As rings ($N_{\rm r}$).  The inset of the (c) 
displays the constructing unit of arsenene. 
The $r_{1}$ and $\theta_{1}$ describe mainly the in-plane bondings.
and the  $r_{2}$ and $\theta_{2}$ represent primarily 
the out-of-plane bondings. 
(d) The total energy changes under straining in 4-aANR 
(black cubics, left scale) and 4-zANR
 (red circles, right scale).   The positive (negative)
  values stand for tensile (compressive) strains, respectively.}

\label{fig1}
\end{figure}

\clearpage

\begin{figure}
\includegraphics[width=16cm]{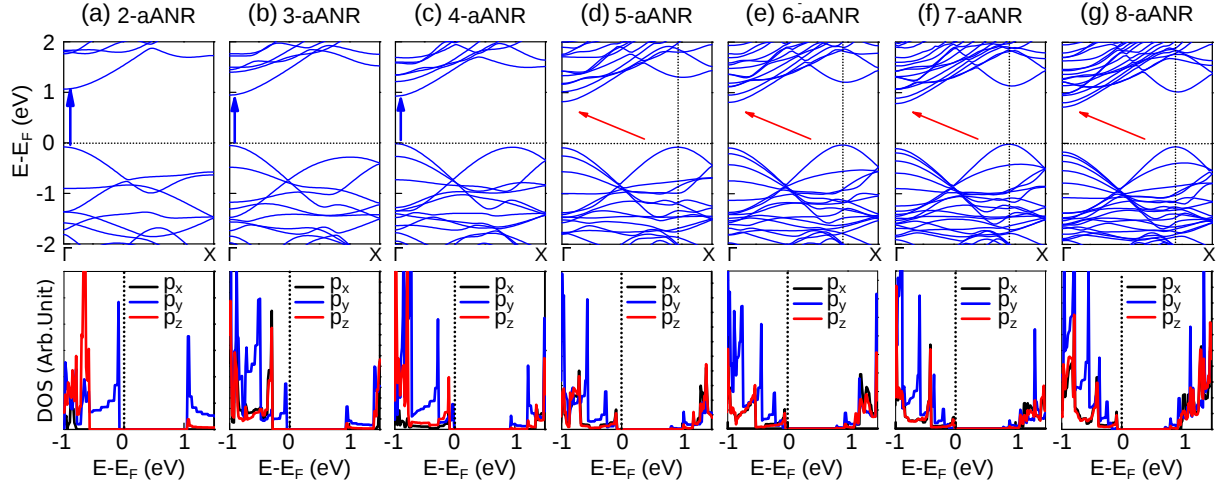}
\caption{(color online) Top panels: band structures for
 $N_{\rm r}$-aANR with (a)  $N_{\rm r}$ = 2, (b) $N_{\rm r}$ = 3,
(c) $N_{\rm r}$ = 4, (d) $N_{\rm r}$ = 5, (e) $N_{\rm r}$ = 6, 
(f) $N_{\rm r}$ = 7 and (g) $N_{\rm r}$ = 8.
 Vertical blue arrows indicate direct bandgaps and
tilted red arrows represent indirect bandgaps.
The valence band maxima for the indirect bandgaps
 are marked with vertical dotted lines.
The horizontal dotted lines represent the
Fermi energy level (E$_{\rm F}$). Bottom panels:
 orbital-resolved density of states (DOS) for the
 corresponding band structures.
 The vertical dotted lines in the DOS plots 
represent the E$_{\rm F}$ levels. }
\label{fig2}
\end{figure}

\begin{figure}
\includegraphics[width=16cm]{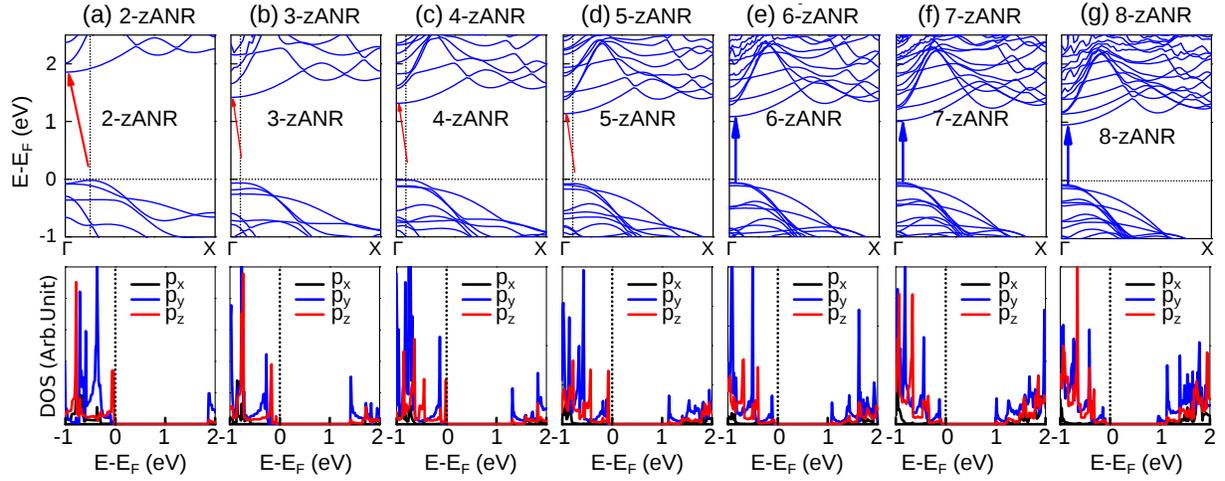}
\caption{(color online)  Top panels: band structures for
 $N_{\rm r}$-zANR with (a)  $N_{\rm r}$ = 2, (b) $N_{\rm r}$ = 3,
(c) $N_{\rm r}$ = 4, (d) $N_{\rm r}$ = 5, (e) $N_{\rm r}$ = 6, 
(f) $N_{\rm r}$ = 7 and (g) $N_{\rm r}$ = 8. 
 Tilted red arrows represent indirect bandgaps and
 vertical blue arrows indicate direct bandgaps.
The valence band maxima for the indirect bandgaps
 are marked with vertical dotted lines.
The horizontal dotted lines represent the
Fermi energy level (E$_{\rm F}$). Bottom panels:
 orbital-resolved density of states (DOS) for the
 corresponding band structures. The vertical
 dotted lines in the DOS plots 
represent the E$_{\rm F}$ levels. }
\label{fig3}
\end{figure}

\clearpage

\begin{figure}
\includegraphics[width=16cm]{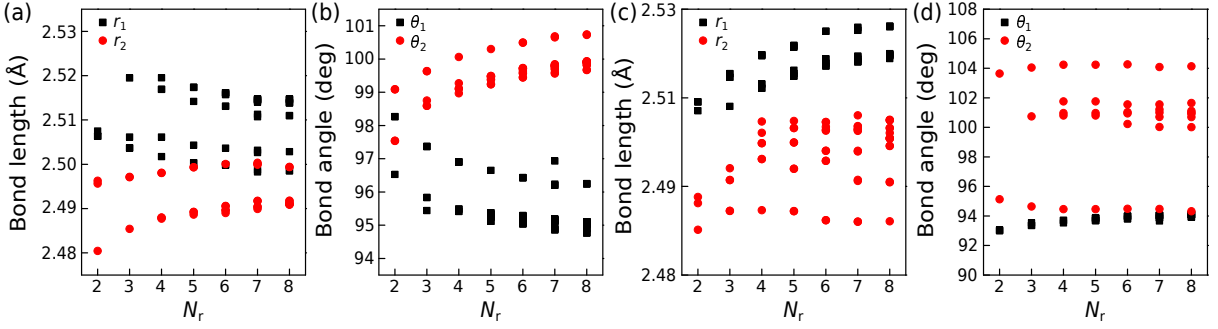}
\caption{(color online) The distributions of 
the bond lengths of $r_{1}$ 
(black cubics) and $r_{2}$ (red circles)
 depending on the number of As-As rings ($N_{\rm r}$) 
 in (a) aANRs and (c) zANRs. 
The changes in the bond angles of $\theta_{1}$ (black cubics)
 and $\theta_{2}$   (red circles) with 
the N$_{\rm r}$ in (b) aANRs and (d) zANRs.}
\label{fig4}
\end{figure}

\clearpage

\begin{figure}
\includegraphics[width=16cm]{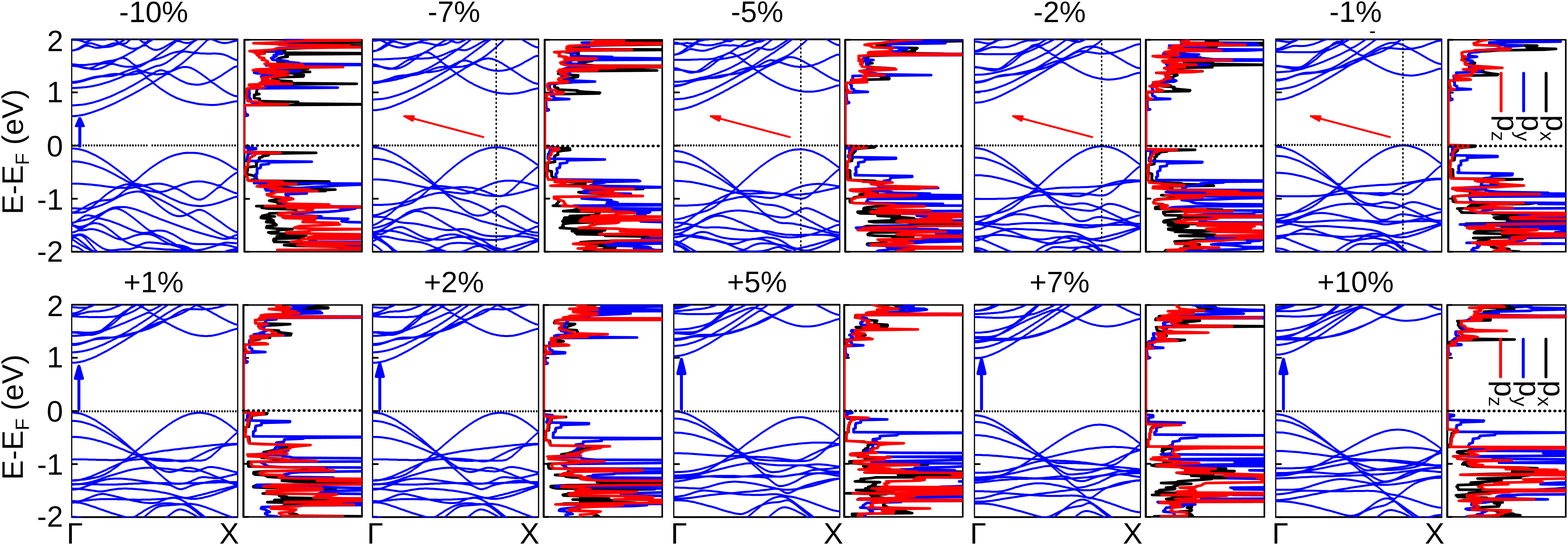}
\caption{(color online) The band structures of 5-aANR at strains  
between $\varepsilon$ = -10\% to $\varepsilon$ = +10\%. 
Vertical blue arrows indicate direct bandgaps.
Tilted red arrows represent indirect bandgaps.
The valence band maxima
 for the indirect bandgaps
 are marked with vertical dotted lines.
The horizontal dotted lines represent the
Fermi energy level (E$_{\rm F}$). The corresponding orbital-resolved
density of states (DOS) are displayed in the right panels for 
each of the band structures. The  dotted lines in the DOS plots 
represent the E$_{\rm F}$ levels.  }
\label{fig5}
\end{figure}

\clearpage

\begin{figure}
\includegraphics[width=16cm]{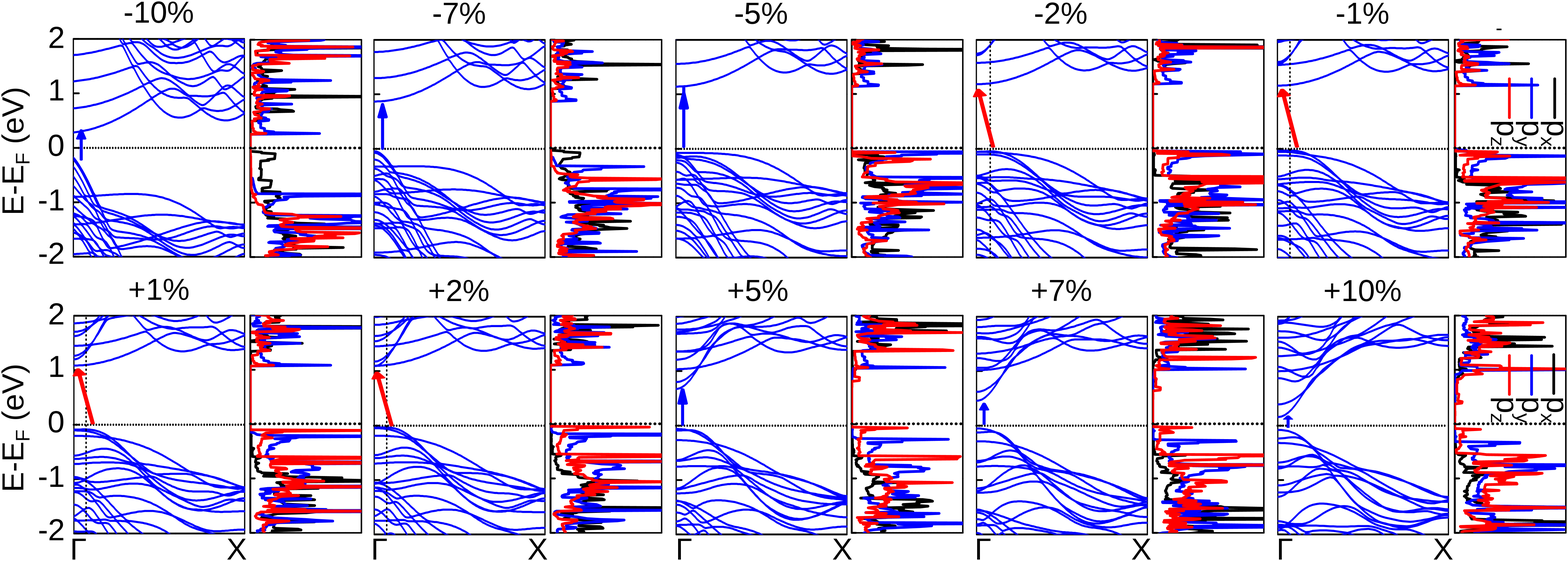}
\caption{(color online) The band structures of 5-zANR at strains 
between  $\varepsilon$ = -10\% to $\varepsilon$ = +10\%. 
Vertical blue arrows indicate direct bandgaps.
Tilted red arrows represent indirect bandgaps.
The valence band maxima
 for the indirect bandgaps
 are marked with vertical dotted lines.
The horizontal dotted lines represent the
Fermi energy level (E$_{\rm F}$). The calculated orbital-resolved
density of states (DOS) are displayed in the right panels for 
the corresponding band structures. The  dotted lines in the DOS plots 
represent the E$_{\rm F}$ levels. }
\label{fig6}
\end{figure}

\clearpage

\begin{figure}
\includegraphics[width=16cm]{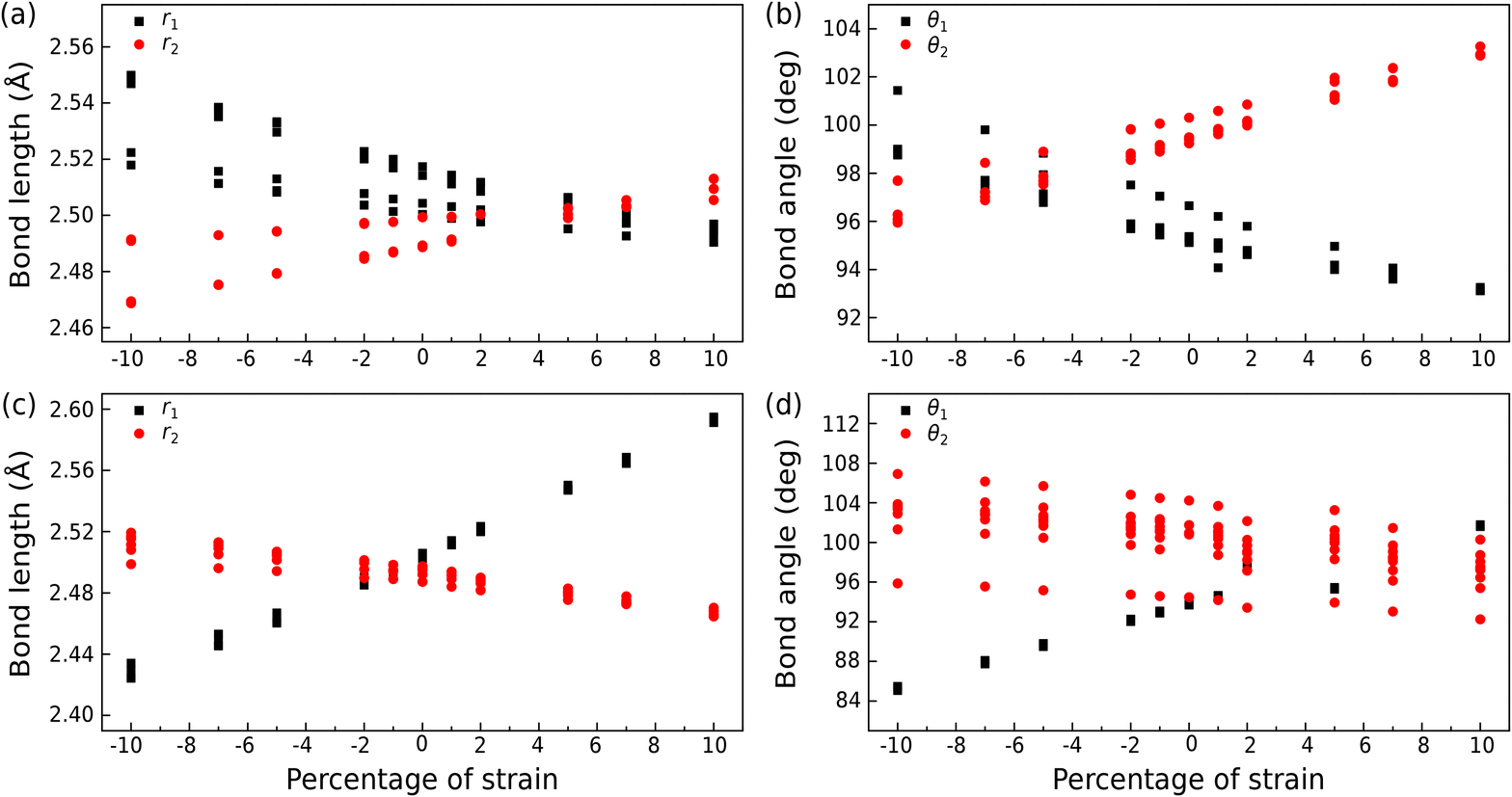}
\caption{(color online) The dependence of bond lengths of $r_{1}$
(black cubics) and  $r_{2}$ (red circles) on various strains in
(a) aANRs and (c) zANRs. The changes of bond angles of
 $\theta_{1}$ (black cubics) and $\theta_{2}$ (red circles) 
 with various strains in (b) aANRs and (d) zANRs.}
\label{fig7}
\end{figure}

\clearpage

\begin{table*}[!hbp]
\caption{Number of As-As rings ($N_{\rm r}$), 
nanoribbon widths ($W_{\rm D}$, in unit of \AA), lattice
  constants ($a$ for aANR and $b$ for zANRs, in unit of \AA) and bandgap
 energy ($E_{\rm g}$, in unit of eV) for aANRs and zANRs.
 The subscript `${\rm d}$'
 marks a direct bandgap and `${\rm i}$'
  denotes an indirect bandgap. }

\begin{tabular}{ccccccc}
\hline
\hline
& aANRs  &  &  &  zANRs  &  & \\
\hline
  $N_{\rm r}$  &  $W_{\rm D}$  &  $a$  & $E_{\rm g}$  &   
  $W_{\rm D}$  &  $b$  & $E_{\rm g}$ \\
\hline
2 &9.5   &4.39  &1.15$^{\rm d}$      &8.9  &3.63 &1.87$^{\rm i}$ \\
3 &13.3  &4.51  &0.99$^{\rm d}$      &13.7 &3.64 &1.48$^{\rm i}$ \\   
4 &16.9  &4.57  &0.94$^{\rm d}$      &18.6 &3.65 &1.33$^{\rm i}$ \\ 
5 &20.6  &4.60  &0.90$^{\rm i}$      &23.4 &3.66 &1.21$^{\rm i}$ \\
6 &24.3  &4.63  &0.85$^{\rm i}$      &28.2 &3.66 &1.15$^{\rm d}$ \\
7 &28.1  &4.65  &0.80$^{\rm i}$      &32.9 &3.66 &1.07$^{\rm d}$ \\
8 &31.7  &4.66  &0.79$^{\rm i}$      &37.7 &3.67 &1.06$^{\rm d}$ \\
\hline
\hline
\end{tabular}
\label{table1}
\end{table*}

\clearpage

\begin{table*}[!hbp]
\caption{Bandgaps and  effective masses of carriers
  with respect to strains in 5-aANR and 5-zANR. The bandgap
  energy is represented by $E_{\rm g}$ (in unit of eV). 
The subscript `${\rm d}$' denotes a direct bandgap and  `${\rm i}$'
  represents an indirect bandgap. The effective masses of holes ($m^{*}_{h}$)  and
 of electrons ($m^{*}_{e}$) are calculated for direct bandgaps and those
  for indirect bandgaps are neglected with `N/A'.}

\begin{tabular}{ccccccc}
\hline
\hline
  & 5-aANR &  &  &  5-zANR &  & \\
\hline
 $\varepsilon$  &  $E_{\rm g}$(eV)  & $m^{*}_{h}/m_{0}$  &  $m^{*}_{e}/m_{0}$  &  $E_{\rm g}$(eV)
 &  $m^{*}_{h}/m_{0}$  &  $m^{*}_{e}/m_{0}$ \\
\hline
+10\%  &1.12$^{\rm d}$  &0.19 &0.21   &0.24$^{\rm d}$   &0.41 &0.11  \\
+7\%   &1.07$^{\rm d}$  &0.20 &0.23   &0.51$^{\rm d}$   &0.53 &0.10  \\   
+5\%   &1.03$^{\rm d}$  &0.20 &0.24   &0.71$^{\rm d}$   &0.65 &0.10  \\ 
+3\%   &0.96$^{\rm d}$  &0.21 &0.26   &1.13$^{\rm i}$   &N/A  &N/A   \\
+1\%   &0.94$^{\rm d}$  &0.22 &0.27   &1.17$^{\rm i}$   &N/A  &N/A   \\
0      &0.90$^{\rm i}$  &N/A   &N/A   &1.21$^{\rm i}$   &N/A  &N/A   \\
-1\%   &0.86$^{\rm i}$  &N/A   &N/A   &1.19$^{\rm i}$   &N/A  &N/A   \\
-3\%   &0.82$^{\rm i}$  &N/A   &N/A   &1.20$^{\rm i}$   &N/A  &N/A   \\   
-5\%   &0.77$^{\rm i}$  &N/A   &N/A   &1.20$^{\rm d}$   &0.13 &1.24  \\
-7\%   &0.70$^{\rm i}$  &N/A   &N/A   &0.92$^{\rm d}$   &0.12 &1.23  \\
-10\%  &0.61$^{\rm i}$  &0.31  &0.42  &0.32$^{\rm d}$   &0.19 &1.20  \\
\hline
\hline
\end{tabular}
\label{table2}
\end{table*}

\end{document}